\documentclass[12pt]{iopart}
\usepackage{graphicx}
\usepackage{bm}
\newcommand{\sech}{\mathop{\mathrm{sech}}}
\begin{document}

\title[Dimerization process and elementary excitations]{Dimerization process and elementary excitations in spin-Peierls chains coupled by frustrated interactions}
\author{D Mastrogiuseppe, C Gazza and A Dobry}
\address{Facultad de Ciencias Exactas Ingenier{\'\i}a y
Agrimensura, Universidad Nacional de Rosario and Instituto de
F\'{\i}sica Rosario, Bv. 27 de Febrero 210 bis, 2000 Rosario,
Argentina.}
\ead{\mailto{dmastro@ifir.edu.ar}}

\begin{abstract}
We consider the ground state and the elementary excitations of an array of spin-Peierls
chains coupled by elastic and magnetic interactions. It is expected that the effect of the
magnetic interchain coupling will be to reduce the dimerization amplitude and that of the
elastic coupling will be to confine the spin one-half solitons corresponding to each isolated
chain. We show that this is the case when these interactions are not frustrated. On the
other hand, in the frustrated case we show that the amplitude of dimerization in the ground
state is independent of the strength of the interchain magnetic interaction in a broad range of
values of this parameter. We also show that free solitons could be the elementary excitations
when only nearest neighbour interactions are considered. The case of an elastic interchain coupling
is analyzed on a general energetic consideration. To study the effect of the magnetic interchain
interaction the problem is simplified to a two-leg ladder which is solved using density matrix renormalization 
group (DMRG) calculations.
We show that the deconfinement mechanism is effective even with a significantly strong antiferromagnetic
interchain coupling.
\end{abstract}


Elementary excitations of antiferromagnetic systems are spin-one
magnons which are bosonic in character and correspond to flipping
locally the spin of one electron. They give rise to a definite
peak in the dynamical structure factor which is observed in
neutron scattering measurements. In a one-dimensional
antiferromagnet, this picture breaks down and the concept of
fractionalized excitation has been proposed in the last years to
explain the excitation spectra. The excitation spectrum of a 1D
antiferromagnetic Heisenberg model is a paradigmatic example of
a fractional quantum state. The elementary excitations are
spin-$\frac{1}{2}$ spinons which are topological excitations identified as
quantum domain walls. The spectral signal corresponding to the excitation of
two spinons shows a highly dispersive continuum
without a definite one-particle peak in the dynamic
susceptibility. Measurements of the two-spinon continuum have been
achieved in the quasi-one-dimensional antiferromagnetic system
KCuF$_{3}$ \cite{tennant}. Since the discovery of the cuprate
superconductors an intense activity has been carried out to see if
a fractional quantum state is possible in a two-dimensional
antiferromagnet. An experimental realization seems to be found in
Cs$_{2}$CuCl$_{4}$ \cite{Coldea}.

Some analogies appear in the excitation spectra when the magnetism
is coupled to phonons in the so-called spin-Peierls
systems. This coupling leads to a magneto-structural transition
towards a dimerized low-temperature phase, opening a gap in the
magnetic spectra. Even though the excitation spectrum of an isolated spin-phonon
chain is not exactly known, an early semiclassical analysis of a
bosonized field theory \cite{nakafuku} gives some insight to this
problem.  The elementary excitation is a soliton, a topological
defect which separates two different phases of the dimerized order.
It is a mixed magnetic and structural excitation carrying a spin-$\frac{1}{2}$.
Therefore, the dynamical response of a pure spin-phonon system should be
 dominated by a two-soliton continuum above the spin gap. Numerical calculations
in small chains confirm this scenario \cite{Augier}.

The role of the interchain coupling should be taken into account for
application to a real compound. It was particularly analyzed in
connection with neutron and optical spectra measurements in the
inorganic spin-Peierls system CuGeO$_3$ \cite{Affleck,dobryiba}.
It has been concluded that solitons are always confined by the interchain
coupling. It means that no trace of solitons is present in the excitation spectrum.
Different routes have been followed to analyze the effect of the
interchain coupling. In one of them a linear confinement potential
between the solitons has been considered as a result of a mean
field treatment of the interchain coupling \cite{Affleck}. A
ladder of bound soliton-antisoliton states appears before a magnon
continuum. In other approach \cite{dobryiba}, a mixed lattice and
magnetic excitation called \emph{domain} was found in a semiclassical
approximation of the bosonized theory of the two-dimensional
spin-phonon system. The domain is a triplet excitation. Excitations
inside the domain give rise to a series of states before the
appearance of a two-domain continuum. Experiments in CuGeO$_3$ \cite{Ain}
have shown the existence of a peak separated from a continuum, in consistency
with the previous pictures. The peak could be associated with a
magnon excitation and the continuum with a two-magnon continuum.

We note that by integrating out the phonon coordinates
the adiabatic-antiadiabatic crossover has been studied in the
one-dimensional model \cite{Citro}.
Moreover, the dynamic correlation function has been studied in the antiadiabatic limit
for the problem of coupled chains \cite{Essler}.
The effective one-chain problem has been solved by using exact results from the sine-Gordon
theory corresponding to the continuum limit of the effective magnetic problem.
The interchain interactions have been taken into account by a RPA approach.
The  dynamic susceptibility has also a peak corresponding to a spin-1 excitation
(which we call a magnon) and a continuum.

Now the question arises about if free solitons could be observed in a real system.
That is to say, if it is possible to observe a two-soliton continuum above the gap
in a spin-Peierls material.
The complete answer to this question is a formidable problem because it implies
the calculation of the dynamic spin-spin correlation
function for a coupled 2D or 3D spin-phonon problem. As a first step we use a semiclassical
approach in the present work. Our purpose is to show that a frustrated nearest neighbour
interchain coupling in a quasi-one-dimensional spin-Peierls system does not confine the in-chain solitons.

Our work is also motivated by the recently identified materials TiOCl and TiOBr as
spin-Peierls compounds \cite{TiOCl}.
These materials deviate from a canonical spin-Peierls behaviour due to the existence
of an intermediate incommensurate phase (between the uniform and dimer phases). It has been shown
that a frustrated interchain coupling in the bi-layer structure could be the origin of that incommensurate phase \cite{incomm}.
Therefore we propose as a minimal model for this material a set of antiferromagnetic spin-Peierls chains
coupled by a frustrated magneto-elastic interaction. The result of the present paper
could be taken as a starting point to interpret the magnetic excitation spectra in the low-temperature
phases of these materials.

Aiming to discuss the previous posed problem, in this work
we compare two different models for the interchain coupling. In
both models an array of one-dimensional Heisenberg chains coupled
to the lattice deformation is described by the following Hamiltonian:
\begin{eqnarray}
 H_{in}&=& \sum_{i,j}\frac{(P_{i}^{j})^2}{2m} +
\frac{1}{2}K_{\mathrm{in}} \sum_{i,j} (u_{i+1}^{j}-u_{i}^{j})^2 \nonumber\\
&+& J_{\mathrm{in}} \sum_{i,j} \left[1+\alpha
(u_{i+1}^{j}-u_{i}^{j})\right]\, \textbf{S}_{i}^{j} \cdot
\textbf{S}_{i+1}^{j}
\label{hchains}
\end{eqnarray}
where we denote by $i$ the site of the $j$th chain, and the 'in' subscript indicates that
this part of the Hamiltonian takes into account only the in-chain interactions.
$\textbf{S}_{i}^{j}$ are spin-$\frac{1}{2}$ operators with exchange
constant $J_{\mathrm{in}}$ along the x-axis of a non-deformed
underlying lattice. In our simplified model we took the scalar ionic coordinates $u^j_i$
to be the relevant for the dimerization process and $P^j_i$ are their conjugate momenta.
$K_{\mathrm{in}}$ is the elastic coupling along the chain. Furthermore $\alpha$ measures the deformation effect
on the magnetic exchange constant.

The two cases we compare are differentiated by the interchain coupling configuration.
In the first one, the magnetic sites lay on a lattice with square geometry. The
interchain coupling reads:

\begin{eqnarray}
H_\mathrm{inter}^\mathrm{sq}&=&\frac{1}{2}K_\mathrm{inter} \sum_{i,j} (u_{i}^{j+1}-u_{i}^{j})^2
+J_\mathrm{inter}\sum_{i,j} \textbf{S}_{i}^{j} \cdot
\textbf{S}_{i}^{j+1}
\label{hintersquare}
\end{eqnarray}

In the other case they reside on an underlying triangular lattice. Equation (\ref{hintersquare}) changes to:
\begin{eqnarray}
H_\mathrm{inter}^\mathrm{tr}&=&\frac{1}{2}K_\mathrm{inter} \sum_{i,j} \left[(u_{i}^{j+1}-u_{i}^{j})^2+(u_{i+1}^{j-1}-u_{i}^{j})^2\right]
\nonumber\\
&+&J_\mathrm{inter}\sum_{i,j} (\textbf{S}_{i}^{j} \cdot\textbf{S}_{i}^{j+1}
+ \textbf{S}_{i}^{j} \cdot\textbf{S}_{i+1}^{j-1})
\label{hintertriang}
\end{eqnarray}
where $K_\mathrm{inter}$ and $J_\mathrm{inter}$ are the elastic and magnetic
interchain coupling constants.
We have considered only nearest neighbour interactions in both cases.
We finally get the complete Hamiltonian as follows:

\begin{eqnarray}
H=H_\mathrm{in}+H_\mathrm{inter}
\label{htot}
\end{eqnarray}
where we must replace the second term of the sum with equation (\ref{hintersquare}) or (\ref{hintertriang})
depending on the case.

\begin{figure}[ht]
\centering
\includegraphics[width=8cm]{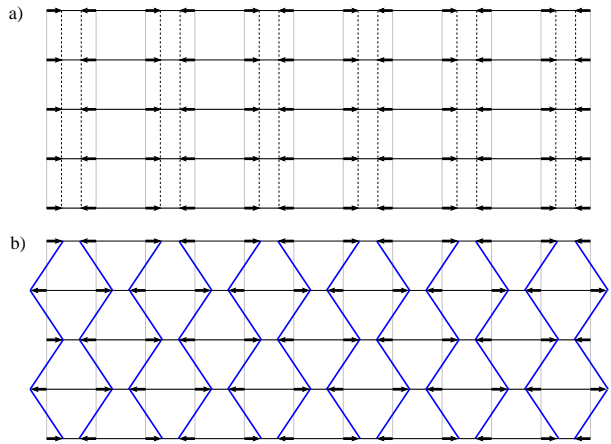}
\caption{\label{dimerlatsq}
Schematic representation of the  dimerized phase in the square lattice (gray background) for
(a) the in-phase and (b) an out-of-phase configuration. Arrows signal the displacement of the atoms. The
thin dashed lines correspond to unperturbed interchain springs and the blue continuous ones forming
zigzag paths are dilated springs.}
\end{figure}

Let us begin with a qualitative energetic interpretation in order to describe
the ground state and the elementary excitations of these systems. We treat the problem
in the adiabatic approximation, i.e. we consider the ionic coordinates as classical static
variables and neglect the kinetic term in (\ref{hchains}).
It is known that the one-dimensional spin-phonon model given by (\ref{hchains}) dimerizes as
the result of a competition between the elastic and magnetic
interactions. In figure \ref{dimerlatsq}  we show the array of dimerized chains (horizontally positioned)
in the square geometry.

Each spin belongs to a singlet and we therefore assume that $J_\mathrm{inter}$ is not active
(at least for a small enough value of $J_\mathrm{inter}$, see the analysis for ladders below).
Therefore, we are not going to consider its effects in the present analysis.
In this figure we represent two possible configurations of dimerized chains. In the upper graphic, each chain is
in phase with its neighboring chains, i.e. the $i$th ion of chain $j$ moves in the same direction of the $i$th ion
in chains $j+1$ and $j-1$. In this case we can observe that this in-phase configuration does not cost additional
interchain elastic energy because every interchain spring is unperturbed. The lower part of the figure shows a possible
out-of-phase configuration where the $i$th ion of chain $j$ moves in opposite direction to that of the $i$th ion
of chains $j+1$ and $j-1$. Here, there is an additional energy cost due to the expansion of every interchain spring.
Any other possible out-of-phase configuration will also have perturbed interchain springs so the in-phase configuration
is the only one that minimizes the total energy.

\begin{figure}[ht]
\centering
\includegraphics[width=8cm]{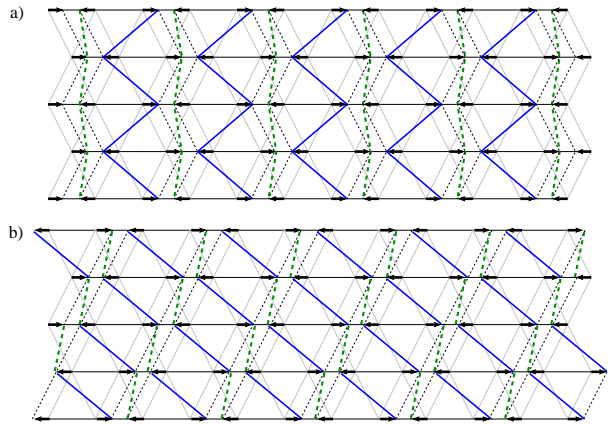}
\caption{\label{dimerlattr}
Schematic representation of the  dimerized phase in the triangular lattice (gray background) for
(a) the in-phase arrangement and (b) an out-of-phase configuration. The thin dashed lines correspond to unperturbed
interchain springs, the thick blue continuous (green dashed) ones forming zigzag paths are dilated (compressed) springs.}
\end{figure}

Now let us see what happens in the triangular geometry. Again, in figure \ref{dimerlattr} we show two possible
arrangements, the in-phase one in the upper part of the figure and an out-of-phase configuration (where
dimers in next-nearest neighbour chains are mutually out of phase) in the lower part.

In this figure we can observe that there is alternately a compressed spring, an unperturbed one, a dilated one and then
another unperturbed spring. Here we have the first difference with respect to the square geometry: meanwhile
in the square geometry there is no additional elastic energy cost in the dimerization process, in the triangular one
there are always deformed interchain springs. The second difference arises when comparing the two panels of figure \ref{dimerlattr}
where we observe that in both cases we have the same deformation pattern of interchain springs, i.e.
due to the frustrated configuration of the model the lower energy state of the system is degenerate with a
degeneracy that depends on all the possible combinations of dimerization patterns. Note that if there were
next-nearest neighbour interactions in the model, the system would lock next-nearest neighbour chains in phase but
there is still no restriction with respect to nearest neighbour chains.

\begin{figure}[ht]
\centering
\includegraphics[width=8cm]{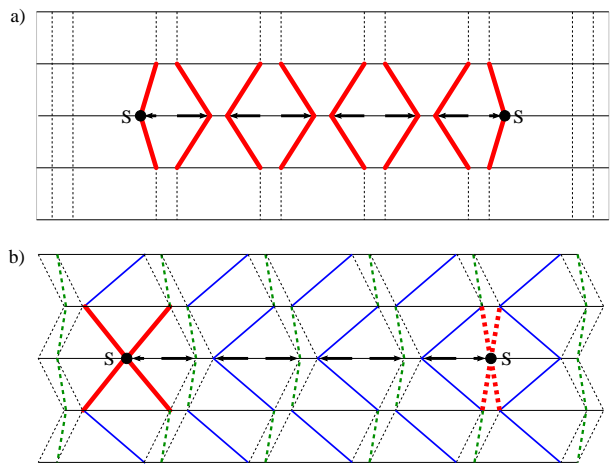}
\caption{\label{2solitint}
Schematic representation of the dimerized phase including two solitons (signalled by an S)
in the central chain for (a) square and (b) triangular lattices. The thick red lines show the perturbation
induced by the excitations on the interchain springs. Note that in (a) every intersoliton spring is affected
and in (b) the excitations produce a local perturbation with respect to the dimerized configuration.}
\end{figure}

Now let us include solitons as point defects dephasing the dimerization order in the in-phase
arrangements of figures \ref{dimerlatsq} and \ref{dimerlattr}. We include two of such defects in the same chain, and
therefore the situation could be represented as shown in figure \ref{2solitint}.
Note the essential difference in both cases, for the square lattice the intermediate zone dimerizes
in antiphase to the background. The energy cost of this configuration is proportional to the distance
between the solitons, so they are always confined with a linear confining potential. In the triangular
geometry this situation does not take place because the interchain energy does not depend on the relative
dimerization phase of two nearest neighbour chains and the only additional energy respect to the
dimerized state is the local cost to create the solitons. This means that the solitons are not
interacting excitations in the triangular geometry.

The previous analysis could be checked quantitatively by the following construction.
As both the lattice dimerization and soliton formation could be taken into account
including a smooth perturbation of the dimerization order, we can approximate
$u_i^j \approx (-1)^\rmi u^j(x)$. The elastic interchain energies of (\ref{hintersquare}) and
(\ref{hintertriang}) could be expanded in a gradient expansion of the lattice constant $a$.

\begin{eqnarray}
H^\mathrm{sq}_\mathrm{el}&=& \frac{K_\mathrm{inter}}{2} \sum_j \int \frac{\rmd x}{a} (u^{j+1}(x)-u^j(x))^2
\label{Helsq}\\
H^\mathrm{tr}_\mathrm{el} &=& K_\mathrm{inter} \sum_j \int \frac{\rmd x}{a} [ 2  u^j(x)^2
+ a\,u^j(x)\partial_x u^{j-1}(x)] \label{Heltriang}
\end{eqnarray}
Note that in the triangular lattice we should retain one higher order in $a$
(the last term of (\ref{Heltriang})) to have a non-null interchain coupling. If it
were not for such term the only effect of the interchain coupling would be to change
the in-chain elastic coupling $K_\mathrm{in}$ by $K_\mathrm{in}+K_\mathrm{inter}$. Therefore the main effect
of the interchain coupling in this case is to weaken the dimerization, reducing the
magnetic gap and increasing the width of the solitons.

As in the previous analysis, a soliton could be included as point defect, thus
neglecting its width. We consider that all but the 0-th chain is dimerized as the
background $u^j(x)=u_0$ for $j \neq 0$. For the 0-th we propose:
\begin{eqnarray}
u_0(x)=u_0 (-1+\Theta(x-x_1))(-1+\Theta(x-x_2))
\end{eqnarray}
being $\Theta(x)$ the Heaviside step function, and where we have put two
solitons at positions $x_1$ and $x_2$ ($x_1 < x_2$). By replacing in
expressions (\ref{Helsq})  we obtain the interacting energy $(4 u_0^2 K_\mathrm{inter}/a)(x_2-x_1)$
which increases linearly with the distance between the solitons. Moreover using
(\ref{Heltriang}) the interacting energy vanishes showing the independence of the solitons.
\newline

In the previous descriptions we have neglected the interchain magnetic interaction. Moreover, we have used a point
approximation for the solitons. In order to undertake a more precise study which also includes the effect of $J_\mathrm{inter}$, we
have to consider the two-chain problem, i.e. we extract a two-leg ladder from the model given in equation (\ref{hchains}),
(\ref{hintersquare}) and (\ref{hintertriang}).
For this purpose, we have performed a numerical analysis for these models, using the DMRG finite algorithm which nowadays is accepted 
as the best numerical tool in low dimensional correlated systems \cite{KarenSchollwock}. 
The ladders were mapped into one-dimensional chains, where interchain interactions become next-nearest neighbour couplings. 
In order to avoid fictitious edge effects, we have performed the calculations with periodic boundary conditions. We have chosen 
100-site ladders (50 sites each leg) to plot our results. We used the soliton width as a parameter to show that the thermodynamic 
limit was reached. Furthermore, this size is long enough to avoid an unwanted overlapping of possibly deconfined solitons due to 
their finite width.
It is worth noting that when the number of sites in each leg is even, as in our case, both solitons will appear in one of
the two chains. On the other hand, with an odd number of sites in each chain, one would obtain one soliton per chain but
the conclusions would remain the same.
We assert the reliability of the calculations keeping 200 states with a truncation error of order $10^{-7}$.
Lattice coordinates were taken into account by solving iteratively the
adiabatic equations, i.e. we take $u^j_i$ as parameters, calculate the
magnetic energy by DMRG, and minimize the total energy (magnetic + elastic)
until convergence. Details of the self-consistent procedure are given
elsewhere \cite{Feiguin, Vekua}.

From now on we set $J_\mathrm{in}$ as the energy scale and $(J_\mathrm{in}/K_\mathrm{in})^{1/2}$ as the displacement one.
We define the dimensionless spin-phonon coupling $\lambda=J_\mathrm{in} \alpha^2/K_\mathrm{in}$. The numerical
results were obtained fixing $\lambda=1$ and changing $K_\mathrm{inter}$ and $J_\mathrm{inter}$.

\begin{figure}[ht]
\centering
\includegraphics*[width=8cm]{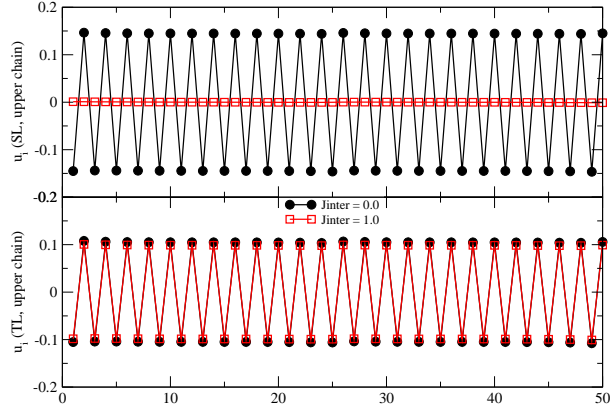}
\caption{\label{ui_sz0}
Lattice distortion in the dimerized state corresponding to the $S_\mathrm{z}=0$ subspace for:
(a) square and (b) triangular ladders. As the distortions in both chains are the same we only show
results for the upper one. We have fixed $K_\mathrm{inter}/K_\mathrm{in}=0.5$.}
\end{figure}

We start by the ground state corresponding to $S_\mathrm{z}=0$. The iterative method converge to the situation
shown in figure \ref{ui_sz0}. Both the square (SL) and the triangular ladders (TL) dimerize, and in the case where there is no magnetic 
interchain coupling, the amplitude of dimerization is smaller in the TL as previously proved.

\begin{figure}[ht]
\centering
\includegraphics*[width=8cm]{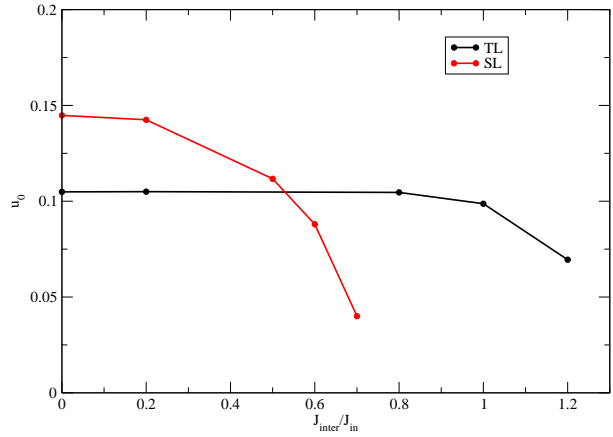}
\caption{\label{dimervsJinter}
Dimerization amplitude as a function of $J_\mathrm{inter}/J_\mathrm{in}$ for the SL and the TL. Again we
use the same elastic parameter indicated in the caption of figure \ref{ui_sz0}.}
\end{figure}

Note that a more essential difference arises if we pay attention to the variation of the dimerization
amplitude $u_0$ with respect to $J_\mathrm{inter}$. The variation of $u_0$ with $J_\mathrm{inter}/J_\mathrm{in}$ is shown in 
figure \ref{dimervsJinter} for both the SL and the TL.

The dimerized state resist a stronger $J_\mathrm{inter}$ for the TL than the SL. Moreover, the dimerization amplitude is independent of $J_\mathrm{inter}$ for the TL in a wide range of this parameter. This is an important fact and implies that in the dimerized state the system remains one-dimensional even with a strong interchain coupling.
A possible interpretation for this result is that each spin is paired in a singlet and in consequence the frustrated interaction cannot compete with this state until it is strong enough to break the singlet and generate an effective uniform 1D Heisenberg model along the zig-zag path.

\begin{figure}[ht]
\centering
\includegraphics*[width=8cm]{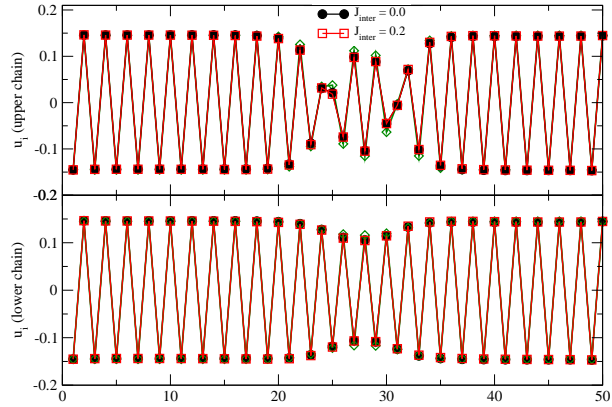}
\caption{\label{usqsz_1} Lattice distortion for the upper and lower chains of the
SL in the $S_\mathrm{z}=1$ subspace. The green curve is a fitting to the analytic result
given by equations (\ref{udomain1}) and (\ref{udomain2}).}
\end{figure}

\begin{figure}[ht]
\centering
\includegraphics*[width=8cm]{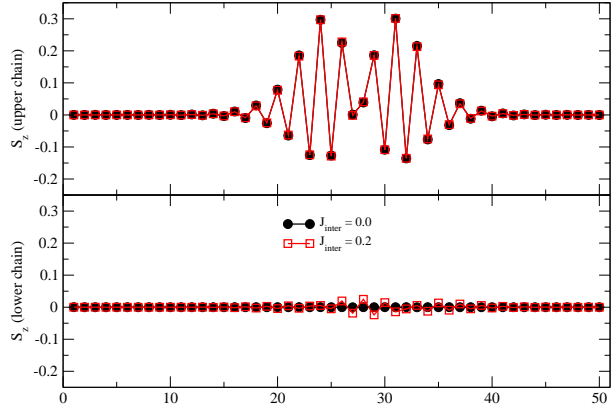}
\caption{\label{szsqsz_1}
Local magnetization for the upper and lower chains of the square ladder
in the subspace $S_\mathrm{z}=1$.}
\end{figure}

Now let us analyze the excited state. In figure \ref{usqsz_1} and \ref{szsqsz_1} we show the lattice deformation
and the local magnetization for the square ladder.

There are two defects of the dimerization pattern. Distortions far away from that defects correspond to
the dimerization pattern with the same value as in  figure \ref{ui_sz0} ($u_0=0.15$). The defect cannot be
considered as two independent solitons because they always fall at the same distance independently of the
initial distortion pattern used in the method. The deformation of the intermediate region between the defects
does not correspond to the dimerization of the bulk and correspondingly $<S_{\mathrm{z}i}>$ does not vanish in this
intermediate zone. Therefore we fail to describe this pattern as a two-soliton form that would be given by
\begin{equation}
    u_i=(-1)^\rmi u_0\, \tanh{\left(\frac{i-x_1}{\xi}\right)} \tanh{\left(\frac{i-x_2}{\xi}\right)}
\label{2solfit}
\end{equation}
where $\xi$ is the soliton width.
It is better described as the domain excitation found in \cite{dobryiba} for the square lattice.
It was obtained by bosonization techniques and the self-consistent harmonic approximation (SCHA).
For comparison, we redo the calculation of \cite{dobryiba} for the two-chain spin-phonon system
coupled by an elastic interaction. To solve the SCHA equations we fix the bosonic field representing
the spin variables in one of the chains (say 1) to its value in the $S_\mathrm{z}=0$ dimerized state \cite{dobryiba}.
The displacement fields $u_1(x)=(-1)^\rmi u^1_i$ and $u_2(x)=(-1)^\rmi u^2_i$ are then given by the following
expressions:
\begin{eqnarray}
u_1(x)=\frac{u_0}{\frac{\epsilon}{2}+1}\left[(1+\frac{\epsilon}{4})+\frac{\epsilon}{4}
t(x-x_\mathrm{M})\right]\label{udomain1}\\
u_2(x)=\frac{u_0}{\frac{\epsilon}{2}+1}\left[\frac{\epsilon}{4}+(1+\frac{\epsilon}{4})
t(x-x_\mathrm{M})\right] \label{udomain2}
\end{eqnarray}
where $\epsilon=K_\mathrm{inter}/K_\mathrm{in}$, $x_\mathrm{M}$ sets the position of the domain in the chain 2 and the function
$t(x)$ is given by:
\begin{eqnarray}
t(x)&=&
1-2 \cosh^2(x_0/\xi) \nonumber\\
&\times& \sech\left[\frac{(x-x_0)}{\xi}\right] \sech\left[\frac{(x+x_0)}{\xi}\right]
\end{eqnarray}
with:
\begin{eqnarray}
\frac{x_0}{\xi}=\frac12 \log\left[\frac{2+B+2\sqrt{1+B}}{B}\right],\hspace{0.5cm}B=\frac{1}{1+\frac{4}{\epsilon}}
\label{x0}
\end{eqnarray}
being $\xi$ a characteristic width of the wall of the domain. $2x_0$ measures the 'radius' of the domain
configuration. Note that for each interchain coupling $\epsilon$, $x_0$ is fixed by equation (\ref{x0}), i.e.
the wall of the domain accommodates at a given equilibrium distance. Only for negligible values of
$\epsilon$ the domain dissociate in two separate domain walls as given in equation \ref{2solfit}.
For any finite $\epsilon$, the walls are as close as they can in order to reduce the interchain energy.

An accurate fitting could be found between the DMRG results and the analytic form given in equation
(\ref{udomain2}). Lines in figure \ref{usqsz_1} were obtained from this equation with $\xi=1.84$ and
$x_\mathrm{M}=27.71$. $x_0/\xi=1.82$ has not been taken as a fitted parameter but obtained from equation (\ref{x0}).
The slight reduction of the dimerization in the lower chain is well predicted by the analytic expression in equation ($\ref{udomain1}$). With the same values for the parameters a very good agreement is found as shown
in the lower panel of figure \ref{usqsz_1}. The overall coincidence prove that the spin-1 excitation we found
numerically could be identified with the domain predicted in analytic calculations based on bosonization.

On the other hand, we also study the effect of a  magnetic interchain coupling ($J_\mathrm{inter}/J_\mathrm{in}=0.2$).
The only perturbation is a small magnetic polarization of the lower chain debilitating the singlets in this chain as observed in figure \ref{szsqsz_1}. The displacements $u_i$ do not change with the presence of $J_\mathrm{inter}$, and the distortions are shown in figure \ref{usqsz_1}.

\begin{figure}[ht]
\centering
\includegraphics*[width=8cm]{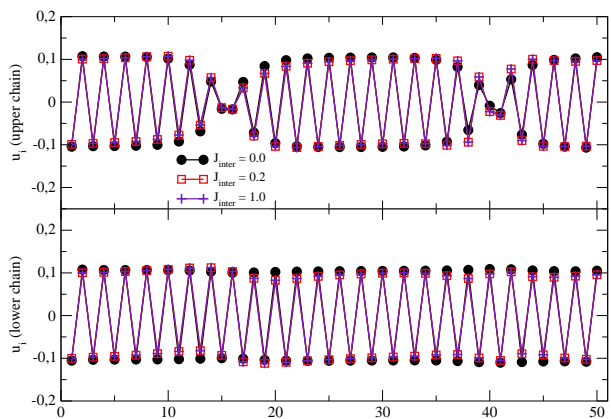}
\caption{\label{uzzsz_1}
Lattice distortion for the upper and lower chains of the TL
in the subspace $S_\mathrm z=1$.}
\end{figure}

A quite different situation occurs in the triangular ladder. In this case the two defects appear
separate enough to give the possibility that the intermediate zone dimerizes as the bulk. The situation
is shown in figure \ref{uzzsz_1}.
The two defects of the dimerization order are now far apart. The
deformation pattern can now be fitted by a two-soliton function as given in equation (\ref{2solfit}).
For the set of parameters we used, we have found a soliton width of $\xi\simeq 3.096$.
Different runs give patterns like those of figure \ref{uzzsz_1} but the distance between the solitons
$x_2-x_1$ depends on the initial distortion pattern. As our method minimizes the total energy this is
an indication that solitons are independent excitations.

\begin{figure}[ht]
\centering
\includegraphics*[width=8cm]{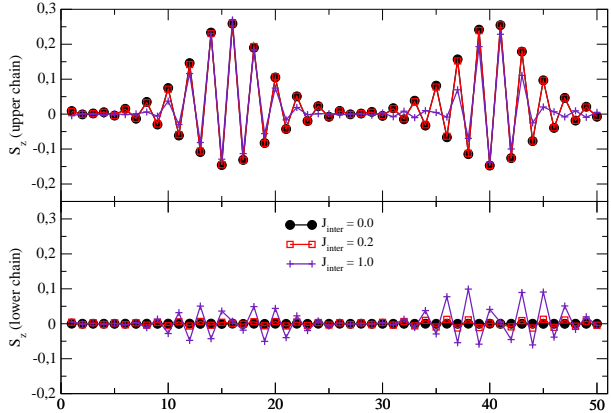}
\caption{\label{szzzsz_1}
Local magnetization for the upper and lower chains of the TL in the subspace $S_\mathrm z=1$.}
\end{figure}

Let us analyze the effect of the magnetic interchain coupling. As a soliton could be thought as a
free spin in a background of singlets it is expected that a weak $J_\mathrm{inter}$ should not have much
influence in the soliton formation and their interactions. This is in fact the case for
$J_\mathrm{inter}/J_\mathrm{in}=0.2$ as shown in figure \ref{szzzsz_1}.
It is interesting to remark that even when the interchain coupling equals the in-chain one the basic
physics does not change. The distortion pattern is described by a two-soliton function and the local
magnetization of the upper chain increase but the perturbation remains localized. Therefore we predict
that independent solitons could be the elementary excitations even in this case.

\begin{figure}[ht]
\centering
\includegraphics*[width=8cm]{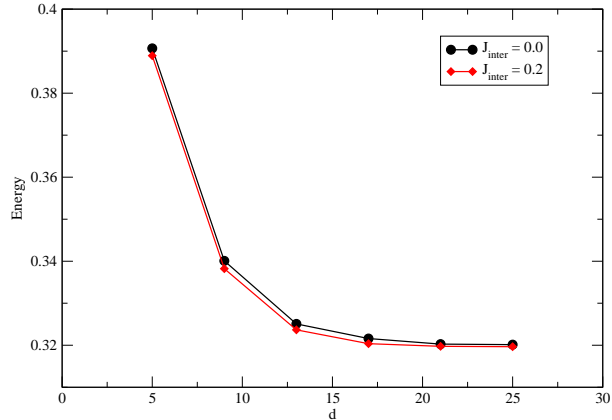}
\caption{\label{eintezz}
Dependence of the interaction energy with the distance between two solitons, for the
triangular ladder.}
\end{figure}

Until now we have found the lattice deformation by minimizing the total energy of the system.
Now we take the solitons as elementary entities and analyze the question of their interaction energy.
Once the distortions have been fitted by the function (\ref{2solfit}) we modify the distance between
the solitons changing the values of $x_2-x_1$. We calculate the total energy as a function of the
distance $d=x_2-x_1$. Results are shown in figure \ref{eintezz}.
We have subtracted the energy of the dimerized state in order to obtain the creation energy of the solitons
plus their interaction energy. The energy increases at small distances. This effect could be interpreted as
a repulsion between the solitons when their distance is smaller than twice their width. For larger separations
the energy does not depend on the distance. This fact shows that solitons are indeed independent when they are
far apart. The inclusion of a $J_\mathrm{inter}=0.2$ does not modify this picture as shown in the figure \ref{eintezz}.
\newline

The role of the interchain coupling frustration in stabilizing 
deconfined spinons was recently discussed
in purely magnetic coupled chains by Nersesyan and Tsvelik
\cite{Tsvelik1}. To make contact with this study we should
consider our spin-phonon model beyond the adiabatic approximation.
Actually the partition function of the model given by equations
(\ref{hchains}) and (\ref{hintertriang}) could be formulated as a
path integral. Moreover, the lattice coordinates $u_i$ could be
integrated out producing a retarded interaction between the spin
variables. In the antiadiabatic limit this interaction becomes
instantaneous. The effective magnetic model will contain, in
addition to the original interchain magnetic interactions,
four-spin exchange interactions as the one included in previous
models leading to deconfined spinons
\cite{Tsvelik2,Starykh,Batista}. It would be very interesting to
study the spin-phonon model by the techniques used in
\cite{Tsvelik2,Starykh} to analyze the appearance of deconfined
spinons when the phonon dynamics is included. On a general ground
we could assert that the deconfined scenario discussed in the
present paper is connected with a local symmetry transformation
that acts on each individual chain changing one of the two
possible dimerized states into the other one. This emergent $Z_2$
gauge symmetry is at the heart of the deconfinement mechanism
found in the magnetic model in \cite{Batista}. Therefore, we
expect the appearance of a deconfined quantum critical point in
this model as found previously in the simplest spin system.

Note  the differences with the model of spin-phonon chains coupled by
non-frustrated interactions. This model could be mapped into an effective Ising model of coupled chains
\cite{Mostovoy}. The mapping takes into account the underlying  {\it global} $Z_2$
symmetry of the dimer order parameter. The symmetry is spontaneously broken in the
low temperature phase and the solitons are confined.
In the frustrated case the dimerization phases between different chains are uncorrelated and 
the mapping of Ref. \cite{Mostovoy} leads to decoupled Ising chains which is a manifestation of the deconfinement
of solitons discussed in the present paper.

Let us finish with the following remark. As previously discussed, in order to fix the dimerization order
in the direction perpendicular to the magnetic chains, we should add a next-nearest neighbour elastic
interaction. Analogously to the argument used for the square geometry, this effect will provide a
confinement mechanism for the solitons. Even though this coupling might be necessary to modelize the
compounds TiOCl and TiOBr, the results of the present paper could be the starting point to describe
the spectrum at not too low temperatures, i.e. above the point where the elastic next-nearest neighbour
energy starts to be washed out by the thermal energy.
\newline

In summary, we have shown an essential difference in the ground state and the elementary excitations of coupled
spin-Peierls chains depending on the nature of the interchain coupling, i.e. with and without frustration. In the
frustrated case we have found that the amplitude of dimerization in the ground state is independent of the
strength of the magnetic interchain coupling in a broad range of values of the parameter $J_\mathrm{inter}$. Moreover, the
analysis of the elementary excitations showed that solitons of the individual chains survive the inclusion of a
nearest-neighbour interchain coupling. As our approach relay in an static approximation of the phonon field we cannot
make precise predictions on the dynamical response of the system. On a general ground we can speculate that a two-soliton
continuum should appear in the frustrated system. This is a different behaviour than that measured in a non-frustrated
spin-Peierls system.

\ack We thank D. Cabra for helpful discussions.
This work was partially supported by  PIP CONICET (Grant 5036).

\end{document}